\documentclass[fleqn,usenatbib]{mnras}

\usepackage{newtxtext,newtxmath}

\usepackage[T1]{fontenc}
\usepackage{ae,aecompl}

\usepackage{graphicx}	
\usepackage{amsmath}	
\usepackage{amssymb}	
\usepackage{todonotes}

\title[Star-disc (mis-)alignment]{Star-disc (mis-)alignment in Rho Oph and Upper Sco: insights from spatially resolved disc systems with K2 rotation periods}

\author[C. L. Davies]{
Claire L. Davies,$^{1}$\thanks{E-mail: cdavies@astro.ex.ac.uk}
\\
$^{1}$Astrophysics Group, School of Physics, University of Exeter, Stocker Road, Exeter, EX4 4QL, UK\\
}

\date{Accepted XXX. Received YYY; in original form ZZZ}

\pubyear{2018}

\begin{document}
\label{firstpage}
\pagerange{\pageref{firstpage}--\pageref{lastpage}}
\maketitle

\begin{abstract}
The discovery of close in, giant planets (hot Jupiters) with orbital angular momentum vectors misaligned with respect to the rotation axis of their host stars presents problems for planet formation theories in which planets form in discs with angular momentum vectors aligned with that of the star. Violent, high eccentricity migration mechanisms purported to elevate planetary orbits above the natal disc plane predict populations of proto-hot Jupiters which have not been observed with \textit{Kepler}. Alternative theories invoking primordial star-disc misalignments have recently received more attention. Here, the relative alignment between stars and their protoplanetary discs is assessed for the first time for a sample of 20 pre-main-sequence stars. Recently published rotation periods derived from high quality, long duration, high cadence \textit{K2} light curves for members of the $\rho$ Ophiuchus and Upper Scorpius star forming regions are matched with high angular resolution observations of spatially resolved discs and projected rotational velocities to determine stellar rotation axis inclination angles which are then compared to the disc inclinations. Ten of the fifteen systems for which the stellar inclination could be estimated are consistent with star-disc alignment while five systems indicate potential misalignments between the star and its disc. The potential for chance misalignment of aligned systems due to projection effects and characteristic measurement uncertainties is also investigated. While the observed frequency of apparent star-disc misalignments could be reproduced by a simulated test population in which $100\%$ of systems are truly aligned, the distribution of the scale of inferred misalignment angles could not.

\end{abstract}

\begin{keywords}
protoplanetary discs -- stars: formation -- stars: rotation -- stars: variables: T Tauri 
\end{keywords}



\section{Introduction}
The traditional Kant-Laplace nebular hypothesis of star and planet formation - based on the close alignment of the solar rotation axis to the planetary orbital axes in the solar system ($\sim7^{\circ}$; \citealt{Beck05}) - predicts that protoplanetary discs form perpendicular to the rotation axis of their natal collapsing cores. Without additional perturbations, the planets that form in such a disc then have orbital angular momentum vectors aligned with the stellar rotation axis. In stark contrast, observations of the Rossiter-McLaughlin effect among transiting exoplanets have uncovered planet orbital axis--stellar rotation axis misalignments: nearly $40\,$per cent of close-in, giant planets (hot Jupiters) display orbital obliquities of up to, and including, $180^{\circ}$ (i.e. anti-alignment; \citealt{Wright11, Campante16}). 

Hot Jupiters are unlikely to have formed in situ \citep{Rafikov06}. Instead, they are understood to have migrated inward from their initial formation location. Kozai oscillations induced by planet-planet interactions \citep{Naoz11}, star-planet interactions \citep{Fabrycky07}, or planet-planet scattering \citep{Rasio96} are widely considered to explain the short period, high obliquity orbits observed. Tidal orbital circularisation and stellar spin-axis reorientation \citep{Winn10, Albrecht12} are additionally invoked to explain why a greater fraction of hot Jupiters are not observed on more oblique orbits. However, high eccentricity migration mechanisms predict a population of proto-hot Jupiters on highly eccentric orbits \citep{Socrates12} which have not been observed by \textit{Kepler} \citep{Dawson15}. 

Alternative, smoother migration mechanisms have also been proposed to explain the existence of high-obliquity hot Jupiters \citep{Goldreich80, Ida08, Bromley11}. In turn, these require primordial star-disc misalignments, potentially caused by gravitational perturbations from stellar companions on binary orbits inclined with respect to the circum-primary disc plane \citep{Batygin12, Spalding15}; interactions between the stellar magnetic field and the inner disc regions \citep[e.g.][]{Foucart11, Lai11}; or turbulence in the natal molecular cloud during gravitational collapse (e.g. \citealt{Bate10}; \citealt{Fielding15}). 
Observational investigations into the presence of star-disc misalignments have so far been limited to debris disc systems in which the host is a main sequence (MS) star. This is primarily due to their relatively close proximity to Earth compared to protoplanetary disc systems meaning that (i) the discs have been easier to spatially resolve and (ii) for two of the closest debris disc hosts, direct estimates of the stellar rotation axis geometry have been made possible using long baseline optical interferometry (e.g. \citealt{leBouquin09, Monnier12}). Both these systems favour star-disc alignment.

\begin{figure}
	\includegraphics[width=0.33\textwidth]{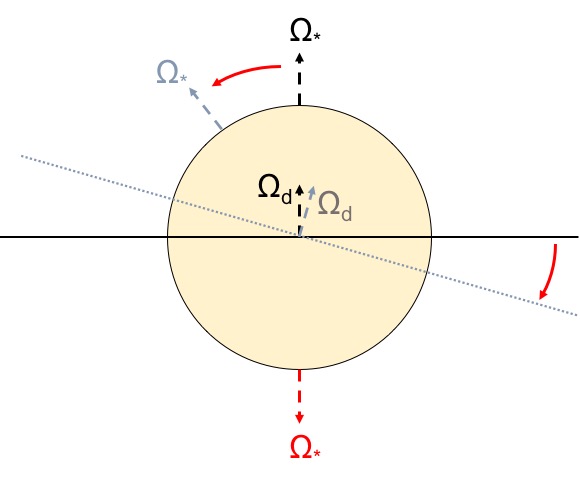}\\
	\includegraphics[width=0.44\textwidth]{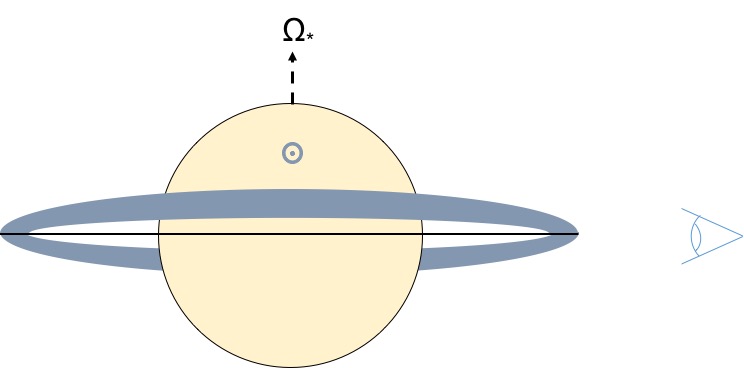}
    \caption{Top: face-on view as seen on-sky; bottom: side-on view of the same star-disc system geometry (the direction to the observer in the bottom figure is to the right, indicated by the eye). The yellow circle represents the star (viewed with angular momentum vector, $\Omega_{\star}$, inclined at $90^{\circ}$) while the black horizontal line represents the disc (viewed with angular momentum vector, $\Omega_{\rm{d}}$, also inclined at $90^{\circ}$). The grey dotted line in the top figure and the corresponding grey torus in the bottom figure illustrate that a disc with non-zero angular momentum \textit{position angle} would still appear aligned with its host star. Similarly, the grey dashed arrow marked $\Omega_{\star}$ in the top figure and the corresponding grey $\odot$ symbol in the bottom figure illustrate that a star with non-zero angular momentum position angle would also appear to be aligned with its disc if just the inclinations are assessed. Finally, the red dashed arrow in the top figure illustrates that an anti-aligned star-disc system ($\Delta i=180^{\circ}$) appear aligned.}
    \label{fig:sketch}
\end{figure}

Statistical assessment of the likelihood of star-disc misalignment in debris disc systems has relied upon more indirect determinations of the stellar inclination, $i_{\star}$ \citep{Watson11, Greaves14}. In these studies, $i_{\star}$ is estimated by combining measurements of the rotation period, $P$, projected rotational velocity, $v\sin i$, and stellar radius, $R_{\star}$, following \citet{Campbell85}: 
\begin{equation}\label{eq:sini}
    \sin i_{\star} = \frac{P}{2\pi}\frac{v\sin i}{R_{\star}}.
\end{equation}
The lack of evidence found for star-disc misalignment in the \citet{Watson11} and \citet{Greaves14} studies has been used to suggest that the occurrence rate of primordial star-disc misalignments is low. However, as noted in \citet{Greaves14}, these studies do not probe the full three dimensional aspect of star-disc alignment (see Fig.~\ref{fig:sketch}) and so misaligned systems may be hidden amongst their samples. Furthermore, this interpretation assumes, for example, that (i) the debris disc traces the same plane as the original protoplanetary disc, (ii) the star-disc obliquity does not alter during pre-main-sequence (PMS) evolution (c.f. \citealt{Rogers12}), and/or that (iii) debris discs form and survive as frequently in misaligned star-disc systems as in aligned systems. Here, the focus of star-disc misalignment is turned to PMS stars. 

\section{Sample selection}\label{sec:sample}

\begin{table*}
	\centering
	\caption{Star and disc parameters compiled from the literature and separated by star forming region (following \citealt{Rebull18}). Column 1: running numerical identifier; column 2: common identifier; column 3: \textit{K2} rotation period; column 4: projected rotational velocity; column 5: spectral type; columns 6 and 7: disc inclination and position angle; column 8: \textit{Gaia} distance; column 9: references for vsini (1: \citealt{Doppmann05}; 2: \citealt{Doppmann03}; 3: \citealt{Kohn16}; 4: \citealt{Dahm12}; 5: \citealt{Chen11}), spectral type (1: \citealt{Erickson11}; 2: \citealt{Wilking05}; 3: \citealt{Prato03}; 4: \citealt{Preibisch99}; 5: \citealt{Cieza10}; 6: \citealt{Preibisch02}; 7: \citealt{Torres06}; 8: \citealt{Preibisch01}; 9: \citealt{Pecaut12}), and disc geometry (1: \citealt{Cox17}; 2: \citealt{Tripathi17}; 3: \citealt{Barenfeld17}; 4: \citealt{Mathews12}; 5: \citealt{Thalmann13}; 6: \citealt{Dipierro18}), respectively; column 10: additional notes (PTD: pre-transition disc; TD: transition disc; B: wide binary; CB: circumbinary disc; DD: debris disc); column 11: $J$-band extinction (see Section~\ref{sec:calcs} for details).}
	\label{tab:lit_values}
	\begin{tabular}{lcccccccccc} 
	    \hline
	    ID & Alt. ID & $P$ & $v\sin i$ & SpT & $i_{\rm{d}}$ & P.A. & Dist. & Refs. & Note & $A_{\rm{J}}$\\
	    & & (days) & (km$\,\rm{s^{-1}}$) &  & ($^{\circ}$) & ($^{\circ}$) & (pc) & & & (mag) \\ 
	    (1) & (2) & (3) & (4) & (5) & (6) & (7) & (8) & (9) & (10) & (11)\\
		\hline
        \multicolumn{10}{c}{$\rho$ Oph} \\
        \hline
        1 & ISO-Oph 51 & 1.8 & $27\pm4.7$ & M$0\pm1$ & $34^{+16}_{-15}$ & $6.1\pm28.3$ & $135.7^{+3.3}_{-3.1}$ & 1, 1, 1 & B & 3.2 \\ 
        2 & WSB 63 & 12.2 & $49\pm7$ & M$1.5\pm0.5$ & $66^{+1.8}_{-1.9}$ & $0.07\pm1.28$ & $137.6\pm2.1$ & 3, 2, 1 & TD & 1.6 \\ 
        3 & DoAr 24E & 5.9 & $39\pm1.9$ & G$6\pm4$ & $53^{+6.2}_{-7.4}$ & $169\pm5$ & $137.3\pm1.7$ & 2, 2, 1 & B & 3.2 \\ 
        4 & Elias 2-24 & 6.6 & $21\pm1.9$ & K$5.5\pm0.5$ & $28.5\pm3.8$ & $46.7\pm6.5$ & $135.7\pm2.0$ & 2, 2, 6 & -- & 3.5 \\ 
        5 & GY 314 & 6.2 & $22\pm1.9$ & K$5\pm0.5$ & $56^{+2.3}_{-2.4}$ & $138.9\pm2.1$ & $136.5^{+1.8}_{-1.7}$ & 2, 2, 1 & PTD & 2.5 \\ 
        6 & DoAr 25 & 8.9 & $10\pm1$ & K$5\pm0.5$ & $63^{+1.9}_{-2.0}$ & $110\pm1.4$ & $137.9^{+1.5}_{-1.4}$ & 3, 2, 1 & TD & 1.0 \\ 
        7 & SR 21A & 2.0 & $65\pm3$ & G$2.5\pm3$ & $18^{+5}_{-9}$ & $10^{+45}_{-49}$ & $137.9\pm1.1$ &  3, 3, 2 & TD;B & 2.7 \\ 
		\hline
        \multicolumn{10}{c}{Upper Sco} \\
        \hline
        8 & USco J160900.0-190836 & 1.8 & $10.41\pm0.30$ & M$5\pm0.3$ & $63^{+18}_{-45}$ & $84^{+81}_{-38}$ & $138.5^{+2.8}_{-2.6}$ & 4, 6, 3 & TD;B & 0.2 \\
        9 & CD-22 11432 & 2.8 & $21.8\pm3.3$ & K$2\pm1$ & $4^{+48}_{-3}$ & $46^{+104}_{-40}$ & -- & 4, 7, 3 & B & -- \\ 
        10 & V935 Sco & 5.0 & $35\pm16$ & K$5\pm2$ & $54^{+4.7}_{-5.5}$ & $46^{+104}_{-40}$ & -- & 3, 5, 1 & CB & -- \\ 
        11 & RX~J1603.9-2031A & 3.9 & $6.37\pm2.91$ & K$5\pm1$ & $69^{+21}_{-27}$ & $5^{+22}_{-26}$ & $142.0\pm0.8$ & 4, 4, 3 & TD;B & 0.5 \\ 
        12 & USco J160357.9-194210 & 3.8 & $11.56\pm0.23$ & M$2\pm0.5$ & $56^{+14}_{-34}$ & $42^{+34}_{-42}$ & $157.3^{+2.1}_{-2.0}$ & 4, 8, 3 & PTD & 0.2 \\
        13 & USco J160643.8-190805 & 7.0 & $13.77\pm2.12$ & K$6\pm0.5$ & $48^{+38}_{-39}$ & $81^{+81}_{-36}$ & -- & 4, 8, 3 & B & -- \\
        14 & USco J160823.2-193001 & 5.4 & $23.98\pm4.27$ & K$9\pm0.5$ & $74^{+5}_{-4}$ & $123^{+3}_{-2}$ & $137.5\pm1.1$ & 4, 8, 3 & PTD;B & 0.5 \\
        15 & USco J160900.7-190852 & 10.0 & $\leq5.7$ & K$9\pm0.5$ & $56\pm5$ & $149\pm9$ & $137.1\pm1.5$ & 4, 8, 3 & PTD;B & 0.2 \\
        16 & USco J160959.4-180009 & 3.7 & $5.89\pm0.76$ & M$4\pm0.5$ & $86^{+4}_{-66}$ & $105^{+59}_{-64}$ & $135.8^{+2.3}_{-2.2}$ & 4, 6, 3 & TD & 0.2 \\
        17 & RX J1614.3-1906 & 4.8 & $17.94\pm1.39$ & K$5\pm0.5$ & $27^{+10}_{-23}$ & $19^{+32}_{-19}$ & $142.5^{+2.6}_{-2.5}$ & 4, 6, 3 & TD & 2.0 \\
        18 & USco J155624.8-222555 & 2.0 & $10.90\pm1.49$ & M$4\pm0.3$ & $85^{+5}_{-67}$ & $53^{+79}_{-27}$ & $140.7^{+2.2}_{-2.1}$ & 4, 6, 3 & TD & 0.1 \\
        19 & RX J1604.3-2130A & 5.0 & $17.3\pm0.4$ & K$2\pm1$ & $6\pm1.5$ & $355\pm10$ & $149.5\pm1.3$ & 4, 4, 4 & PTD;B & 1.0 \\
        20 & HIP 79977 & 1.7 & $57\pm4$ & F$3\pm1$ & $84^{+2}_{-3}$ & $114\pm0.3$ & $131.0\pm0.9$ & 5, 9, 5 & DD & 0.1 \\ 
		\hline
	\end{tabular}
\end{table*}

While limited to just one dimension of a three dimensional problem, comparing star and disc inclinations is still a useful tool to probe star-disc misalignment. The sample considered here is based on matching recently published rotation periods for members of $\rho$~Ophiuchus and Upper Scorpius ($\rho$~Oph and Upper Sco; \citealt{Rebull16}), obtained during NASA's \textit{K2} mission \citep{Howell14}, to previously published spatially resolved disc detections and $v\sin i$ determinations. Only \textit{K2} rotation periods are used as the high quality, long-duration, high cadence light curves (LCs) enable stellar rotation signatures to be identified among other sources of temporal variability exhibited by PMS stars. The uninterrupted coverage afforded by space-based observations also avoids the $1\,$day beat period affecting ground-based period determinations \citep{Davies14}. Space-based rotation periods have also been determined for members of the star forming region NGC~2264 with CoRoT \citep{Affer13} but this region is not considered here as, at $\sim800\,$pc, it is too distant for the discs of its members to be spatially resolved by current interferometric facilities. 

\begin{table*}
	\centering
	\caption{Previously identified members of binary systems in order of binary separation. Column 1: numerical identifier as in Table~\ref{tab:lit_values}; column 2: SIMBAD identifier for companion, where available; columns 3, 4, and 5: projected separation, binary position angle, and magnitude difference in $K$-band reported in the references in column 6 (1: \citet{Chelli88}; 2: \citealt{Reipurth93}; 3: \citealt{Simon95}; 4: \citealt{Kohler00}; 5:  \citet{Prato03}; 6: \citealt{Duchene07}; 7: \citealt{Kraus09}; 8: \citealt{Metchev09}; 9: \citealt{Lafreniere14}; 10: \citealt{Ruiz16}). Column 7, 8 and 9: \textit{Gaia} DR2 parallax and proper motion; column 10: primary (P) or secondary (S) component (except for V935~Sco where the name refers to both components).}
	\label{tab:bin}
	\begin{tabular}{llcccccccc} 
		\hline
        ID & Companion ID & Sep. & P.A. & $\Delta K$ & Refs & $\omega$ & $\mu_{\rm{\alpha}}$ & $\mu_{\rm{\delta}}$ & Note \\
         & & ('') & ($^{\circ}$) & (mag) & & (mas) & (mas/yr) & (mas/yr) & \\
        (1) & (2) & (3) & (4) & (5) & (6) & (7) & (8) & (9) & (10)\\
		\hline
        10 & -- & $0.02029$ & $351.02$ & $0.42$ & 10 & $6.97\pm0.05$ & $-6.93\pm0.11$ & $-26.28\pm0.08$ & -- \\
        9 & -- & $0.222$ & $304.76$ & $0.21$ & 8 & $-4.37\pm0.73$ & $-5.70\pm1.52$ & $-32.63\pm1.03$ & P \\
        13 & -- & $0.247$ & $222.68$ & $1.70$ & 9 & $6.93\pm0.32$ & $-7.06\pm0.65$ & $-19.22\pm0.48$ & P \\
        3 & -- & $2.060$ & $151$ & $1.0$ & 1, 3 & $7.25\pm0.09$ & $-6.86\pm0.23$ & $-26.45\pm0.13$ & P \\
        -- & -- & -- & -- & -- & -- & $7.96\pm0.28$ & $-10.45\pm0.55$ & $-25.20\pm0.31$ & S\\
        7 & -- & $6.4$ & $175$ & $3.30$ & 2, 5 & $7.23\pm0.06$ & $-5.61\pm0.14$ & $-29.08\pm0.09$ & P \\
        -- & EM* SR 21S & -- & -- & -- & -- & $7.16\pm0.27$ & $-6.21\pm0.58$ & $-30.07\pm0.39$ & S \\
        1 & -- & $9.08$ & $152.5$ & $0.37$ & 6 & $7.35\pm0.17$ & $-5.17\pm0.38$ & $-27.46\pm0.23$ & P \\
        -- & 2MASS J16263713-2415599 & -- & -- & -- & -- & $7.47\pm0.14$ & $-6.38\pm0.33$ & $-26.94\pm0.21$ & S \\
        14 & -- & $13.47$ & $71.4$ & $0.41$ & 7 & $7.25\pm0.06$ & $-12.70\pm0.12$ & $-22.26\pm0.09$ & S \\
        -- & USco J160822.4-193004 & -- & -- & -- & -- & $7.15\pm0.05$ & $-12.27\pm0.11$ & $-21.78\pm0.08$ & P \\
        15 & -- & $18.92$ & $326.5$ & $1.81$ & 7 & $7.27\pm0.08$ & $-9.29\pm0.14$ & $-24.92\pm0.09$ & P \\
        8 & USco J160900.0-190836 & -- & -- & -- & -- & $7.19\pm0.14$ & $-10.04\pm0.23$ & $-24.81\pm0.16$ & S \\
        19 & -- & $16.22$ & $215.9$ & $0.92$ & 4, 7 & $6.66\pm0.06$ & $-12.33\pm0.10$ & $-23.83\pm0.05$ & P \\
        -- & RX J1604.3-2130B & -- & -- & -- & -- & $6.79\pm0.10$ & $-12.64\pm0.18$ & $-24.73\pm0.09$ & S \\
		\hline
	\end{tabular}
\end{table*}

From the initial \citet{Rebull18} sample, only stars with an infrared excess (IRex) and a single rotation period measurement were retained. The latter selection criteria avoided including stars exhibiting differential rotation \citep{Rebull16b}. Further requiring that stars were identifiable within SIMBAD for cross-matching purposes provided a starting sample of $200$. The literature was searched for spatially resolved disc detections with $48$ star-disc systems retained at this stage\footnote{EPIC\,203969721 was not retained as it was unclear which component of this multiple system the rotation period belonged to and whether the same component hosts the disc resolved by \citet{Cox17}.}. A further literature search for $v\sin i$ measurements resulted in a sample of $20$ systems (see Table~\ref{tab:lit_values}). 

Measuring $v\sin i$ is not without its difficulties. Acquiring high resolution spectra with a good signal-to-noise ratio (SNR) across photospheric lines is essential. Lines sensitive to other broadening mechanisms (thermal, pressure, and Zeeman broadening, for instance) must ideally be avoided while instrumental broadening should be accounted for in the models of zero rotation. The $v\sin i$ measurements collated here have been determined via different procedures involving the comparison of optical \citep{Chen11,Dahm12,Kohn16} and near-infrared \citep{Doppmann03,Doppmann05} spectra to artificially rotationally-broadened spectral templates \citep{Doppmann03, Doppmann05, Chen11, Kohn16} or slowly rotating ``standard'' MS stellar spectra \citep{Dahm12}. In the latter case, it is assumed that the pressure broadening induced by the lower surface gravities of PMS stars compared to their MS counterparts is minimal.

The \citet{Doppmann05}, \citet{Dahm12}, and \citet{Kohn16} studies explicitly restrict their analysis to lines unaffected by Zeeman broadening. \citet{Chen11} do not make explicit mention of whether the effects of Zeeman broadening are taken into account in their analysis. However, $v\sin i$ measurements are only sourced from this latter study for the debris disc-hosting star, HIP~79977. The location of this object in the Hertzsprung-Russell diagram suggests that the large-scale magnetic field is likely to be low \citep{Gregory12}. As such, Zeeman broadening effects are expected to be minimal.

The $v\sin i$ measurements for DoAr~24E, Elias~2-24, and GY~314, retrieved from \citet{Doppmann03}, were made using spectral lines in the Na interval. \citet{Doppmann03} note these lines may be susceptible to Zeeman broadening effects: large-scale magnetic field strengths of $\sim2\,$kG -- typical for disk-hosting PMS stars (e.g. \citealt{Donati11, Donati12, Donati13}) -- may lead to a $10\%$ over-estimation in their $v\sin i$ measurements. For GY~314, an additional $v\sin i$ estimate was available in the literature: \citet{Greene97} reported $12\pm5\,\rm{km\,s^{-1}}$, lower than that reported by \citet{Doppmann03}. \citet{Greene97} based their estimate on synthetically broadened spectra of a K5 MS star while their spectra were of poorer resolution. The \citet{Doppmann03} $v\sin i$ estimate is adopted herein. The impacts of this are addressed further in Section~\ref{sec:systematics}.

The only star for which an independent estimates of $v\sin i$ had been obtained using sufficiently high-resolution spectra is SR~21A: \citet{James16} found $v\sin i=60\pm5\,\rm{km\,s^{-1}}$, consistent with the estimate adopted here. For all other systems, inspecting the reliability of each $v\sin i$ measurement directly was not possible. However, both \citet{Dahm12} and \citet{Kohn16} cross-checked their $v\sin i$ measurements against previously published values for the $15$ and $2$ stars, respectively, which overlapped with prior high resolution spectroscopic surveys. For all but the most rapidly rotating stars ($v\sin i$ on the order of hundreds of km\,s$^{-1}$), good agreements was found between $v\sin i$ estimates.

\section{Calculation of stellar inclination}\label{sec:calcs}
Determining $i_{\star}$ using equation~(\ref{eq:sini}) requires an estimate of the stellar radius, $R_{\star}$. While it is possible to fit templates to spectra in order to determine $R_{\star}$ directly, spectra with consistent resolution were not available across the full sample of $20$ stars. Model-independent assessments of $R_{\star}$ were instead made from the stellar effective temperature, $T_{\rm{eff}}$, and bolometric luminosity, $L_{\star}$. In turn, estimates of $T_{\rm{eff}}$ were made using spectral-type-to-$T_{\rm{eff}}$ conversions empirically-derived for $5-30\,$Myr old stars \citep{Pecaut13}. These account for the larger surface gravities and spotted surfaces of PMS stars. 

For most of the systems considered here, either a single estimate of the spectral type exists in the literature or multiple studies report consistent spectral type estimates. However, some systems have a broader range of published spectral type estimates. For instance, those for Elias~2-24 range from M2 \citep{Doppmann03} to K5.5 \citep{Wilking05} while \citet{Rigliaco16} report $T_{\rm{eff}}\approx4500\,$K, corresponding to a spectral type $\sim$K3.5 (according to the \citealt{Pecaut13} scale). \citet{Wilking05} and \citet{Gatti06} both report spectral types of K5 for GY~314 while \citet{Luhman99} and \citet{Prato07} report M0 for the same object. Spectral typing SR~21A is complicated by the presence of absorption features across hydrogen emission lines (see discussions in \citealt{Prato03} and \citealt{Kohn16}) with G3 \citep{Prato03} and F7 \citep{Herczeg14} spectral types both reported. In all cases, the spectral types in Table~\ref{tab:lit_values} are chosen as the most commonly-adopted values in the literature.

Values of $L_{\star}$ were calculated by applying a bolometric correction (BC) to a distance modulus- and extinction-corrected 2MASS $J$-band magnitude. A bolometric magnitude for the Sun, $M_{\rm{bol,\odot}}=4.74\,$mag was adopted following \citet{Mamajek12}. Extinction is best assessed using stellar photosphere-tracing spectra. However, only SR~21A, DoAr~24E and USco~J160900.7-190852 had spectroscopically-determined extinction estimates available in the literature \citep{Prato03, Herczeg14, Ansdell16}. Instead, the extinction at $J$-band, $A_{\rm{J}}$, was determined from photometry using
\begin{equation}\label{eqn:av}
    A_{\rm{J}}=2.77[(J-H)-(J-H)_{0}],
\end{equation}
where $(J-H)_{0}$ is the spectral type-dependent intrinsic colour and the value of the constant is valid for the 2MASS photometric system \citep{Yuan13}. The resulting estimates of $A_{\rm{J}}$ are listed in Table~\ref{tab:lit_values}.

This method of estimating the extinction assumes that the stellar photosphere is the only contributor to the observed $J$- and $H$-band flux. In their assessment of the multi-wavelength photometry obtained for the systems in Table~\ref{tab:lit_values}, \citet{Rebull18} identify the shortest wavelength at which the infrared excess becomes significant. The stars in Table~\ref{tab:lit_values} are each reported to display clear infrared excesses emerging longward of $\sim1.6\,\mu$m. However, \citet{Fischer11} showed that accretion disk-hosting, low mass PMS stars may exhibit emission in excess of that expected of a bare stellar photosphere even across $IYJ$ wavebands. To minimise contamination from accretion and thermal disc emission, \citet{Fischer11} recommend using $I$-$J$ colours to determine extinction towards low mass stars rather than $V$-$R$ or $R$-$I$. However, contemporaneous photometry at $I$ and $J$ bands is not available for the systems in Table~\ref{tab:lit_values}. The intrinsically variable nature of PMS stars means that using non-contemporaneous photometry would adversely affect the measurement of the observed colour. As 2MASS photometry is contemporaneous and \citet{Fischer11} highlight that the median level of excess is similar across $J$ and $H$ bands (thus having a minimal effect on the inferred colour), the 2MASS $J$-$H$ colour is used in the estimation of the extinction.

For each star, the 2MASS $J$-band BC and $(J-H)_{0}$ were interpolated from table 6 of \citet{Pecaut13}. The measurement uncertainties on $R_{\star}$ were propagated from the uncertainties in spectral type classification and assuming an error in $\log(L_{\star}/L_{\odot})$ of $0.1\,$dex. This method of calculating $L_{\star}$ assumes that the luminosity spreads observed across individual star forming regions are indicative of a true spread in $R_{\star}$ (cf. discussion in \citealt{Soderblom14}). Although unresolved binaries, differential extinction, photometric variability, and variable accretion all contribute to the observed luminosity spread, they cannot explain its full extent \citep{Hartmann2001, Burningham05, Preibisch05, Slesnick2008}. In addition, $R_{\star}$ spreads have been found independently of $L_{\star}$ spreads \citep{Jeffries07}. 

Two methods are adopted for estimating the distance modulus ($DM$). In the first instance, typical distances to the star forming regions are assigned to all members of each star forming region: $137.3\pm1.2\,$pc for $\rho$~Oph, \citep{OrtizLeon17}, and $145\pm2\,$pc for Upper Sco \citep{deZeeuw99}. Secondly, distance estimates computed from individual \textit{Gaia} DR2 parallax measurements \citep{Gaia16, Gaia18} were retrieved from \citet{Bailer18}. 

\subsection{Accounting for multiplicity}\label{sec:multiplicity}
Of the $20$ star-disc systems in Table~\ref{tab:lit_values}, $10$ exist in confirmed or candidate binary systems. Their projected separations, position angles, $K$-band magnitude difference ($\Delta K$) as well as their Gaia parallax ($\omega$), and proper motions ($\mu_{\alpha}$ and $\mu_{\delta}$) are presented in Table~\ref{tab:bin}. According to the projected binary separation of CD-22~11432 reported in \citet{Metchev09} and the disc radius reported in \citet{Barenfeld17}, the binary companion is located exterior to the outer disc edge. Thus, the disc in this system is not considered to be circum-binary as was assumed in \citet{Kraus12}. 

Additional motion caused by binary companions is not accounted for in the \textit{Gaia} DR2 parallax estimation \citep{Luri18}. The binaries V935~Sco, CD-22~11432 and USco~J160643.8-190805 appear unresolved in \textit{Gaia} DR2. Thus, distance estimates are not provided for these three systems in Table~\ref{tab:lit_values}. The astrometric excess noise values from the \textit{Gaia} astrometric fits were inspected to ensure that the impact of additional motion induced by the presence of companions was minimal (cf. \citealt{Lindegren16}). With the exception of Usco~J160643.8-190805, all astrometric excess noise values were $<0.8\,$milliarcseconds (mas). Individual distances obtained from \citet{Bailer18} for the remaining 17 stars are listed in Table~\ref{tab:lit_values}. 

Flux contamination can bias the stellar inclination estimates if a companion or neighbouring star is sufficiently close. For instance, essentially all of the flux from neighbouring stars with projected separations $\lesssim2$'' will be captured in the 2MASS point spread function (PSF). The three systems which lack individual \textit{Gaia} DR2 parallax measurements lie within this limit. Individual component photometry at $J$- and $H$-bands is not available in the literature for these objects. Consequently, their $R_{\star}$ cannot be determined and they are excluded from further analysis. Another close binary system, DoAr~24E, has a projected separation of $2.060$'' \citep{Simon95}. At this separation, the 2MASS photometry is likely affected by partial Malmquist bias whereby part of the secondary flux is captured within the 2MASS PSF. Individual component $J$- and $H$-band photometry for this binary system, retrieved from \citet{Prato03}, were converted from the CIT to the 2MASS photometric system using the relations in \citet{Carpenter01} and adopted instead.

K2 photometry is also susceptible to contamination by close neighbours. DoAr~24E and SR~21A both have relatively close neighbours but these are both infrared companions \citep{Prato03}. Contamination of their optical LCs is thus unlikely. By comparison, the companion to ISO-Oph~51 is relatively distant. However, the results for this object from K2 appear confused. \citet{Cody17} report that the K2 photometry of ISO-Oph~51 is contaminated by a neighbouring source at $\sim8$''. SIMBAD and \textit{Gaia} DR2 were used to check for neighbouring stars: only 2MASS~J16263713-2415599 is known to exist within $10$'' of ISO-Oph~51 (see Table~\ref{tab:bin}) and is assumed herein to be the likely source of this contamination due to its similar optical brightness\footnote{Individual component photometry in \cite{Wilking05} and \citet{Erickson11} reveal ISO-Oph~51 and 2MASS~J16263713-2415599 have Cousins-$R_{\rm{c}}$ magnitudes of 13.97 and 13.30, respectively.}. In accounting for this contamination, \cite{Cody17} do not recover the $\sim1.8\,$day periodicity reported by \citet{Rebull18}. Instead, \citet{Cody17} find a $\sim20\,$day periodicity in bursting events exhibited in the LC of ISO-Oph~51. This periodicity is not likely to trace stellar rotation. Meanwhile, \citet{Rebull18} also highlight the bursting events within the LC of ISO-Oph~51 while reporting a $\sim20\,$day periodicity in the ``burster'' LC of 2MASS~J16263713-2415599. In light of this apparent confusion, ISO~Oph~51 is removed from further analysis.

As an additional check for potential photometry contamination, a companion search within $10''$ of each of the remaining objects in Table~\ref{tab:lit_values} was performed using the \textit{Gaia} archive\footnote{\url{https://gea.esac.esa.int/archive/}}. No evidence for additional co-distant and co-moving companions was found.

\section{Results and discussion}\label{sec:results}
\begin{table}
	\centering
	\caption{Stellar radius and inclination estimates determined using distance moduli ($DM$) calculated from (i) distances of $137.3\pm1.2\,$pc or $145\pm2\,$pc depending on the star forming region (SFR) membership (columns 2 and 3) and (ii) \textit{Gaia} parallaxes (columns 4 and 5).}
	\label{tab:inc}
	\begin{tabular}{lcccc} 
		\hline
		& \multicolumn{2}{c}{SFR $DM$} & \multicolumn{2}{c}{\textit{Gaia} $DM$}\\
		\hline
		ID & $R_{\star}$ & $i_{\star}$ & $R_{\star}$ & $i_{\star}$\\
		 & ($R_{\odot}$) & ($^{\circ}$) & ($R_{\odot}$) & ($^{\circ}$)\\
		(1) & (2) & (3)  & (4) & (5) \\
		\hline
WSB 63	    & $2.2\pm0.3$ & --	&	$2.2\pm0.3$	& --	\\
DoAr 24E	& $4.8\pm1.0$ & $74\pm6$	& $4.8\pm1.0$ & $74\pm6$ \\
Elias 2-24	& $5.8\pm0.7$ & $28\pm1$	&	$5.7\pm0.7$	&	$29\pm1$	\\
GY 314	&	$2.6\pm0.3$ & $>56$	&	$2.5\pm0.3$	&	$>57$	\\
DoAr 25	&	$2.4\pm0.3$ & $48\pm1$	&	$2.4\pm0.3$	&	$48\pm1$	\\
SR 21A	&	$4.5\pm0.8$ & $35\pm1$	&	$4.5\pm0.8$	&	$35\pm1$\\	
J160900.0-190836	&	$0.9\pm0.1$	&	$24\pm1$	&	$0.9\pm0.1$	&	$25\pm1$	\\
RX J1603.9-2031A	&	$1.9\pm0.3$	&	$15\pm1$	&	$1.8\pm0.3$	&	$16\pm1$	\\
J160357.9-194210	&	$0.9\pm0.1$	&	$71\pm3$	&	$1.0\pm0.1$	&	$61\pm2$	\\
J160823.2-193001	&	$1.3\pm0.2$	& $>35$ &	$1.2\pm0.1$	& $>37$	\\
J160900.7-190852	&	$<1.3$	    &	$<59$	    &	$<1.2$	    &	$<65$	\\
J160959.4-180009	&	$1.1\pm0.2$	&	$24\pm1$	&	$1.0\pm0.1$	&	$26\pm1$	\\
RX J1614.3-1906	& $2.7\pm0.3$ & $39\pm1$ & $2.7\pm0.4$ & $40\pm1$	\\
J155624.8-222555	&	$0.8\pm0.1$	&	$32\pm1$	&	$0.8\pm0.1$	&	$34\pm1$	\\
RX J1604.3-2130A	&	$1.6\pm0.2$	&	$>61$	&	$1.6\pm0.2$	&	$>58$	\\		
HIP 79977  &   $1.8\pm0.2$ &   $>78$   &   $1.6\pm0.2$ &   -- \\
        \hline
	\end{tabular}
\end{table}

$R_{\star}$ and $i_{\star}$ values computed using (i) distances of $137.3\pm1.2\,$pc or $145\pm2\,$ depending on the star forming region (SFR) membership and (ii) \textit{Gaia} parallaxes are presented in Table~\ref{tab:inc}. The two $R_{\star}$ estimates for each star agree to within the derived uncertainties. An estimate of $i_{\star}$ could not be made for five stars as $\sin i_{\star}$ exceeded a value of $1$, regardless of the adopted $DM$. Of these five stars, it is possible that WSB~63 is affected by systematic errors as both $\sin i$ estimates exceeded a value of $1$ by $>1\sigma$. This result is robust against the choice of spectral type for this object\footnote{Alternative spectral types reported in the literature include M$2\pm1$ \citep{Cieza10} and M$0.3\pm0.6$ \citep{Kohn16}.}. Consulting Table~\ref{tab:lit_values}, one can see that this object features both the slowest rotation period and the largest $v\sin i$ of the sample. The period reported for this star may not trace stellar rotation. Instead, it may be associated with a different variability mechanism. 

\begin{figure}
	\includegraphics[width=0.47\textwidth]{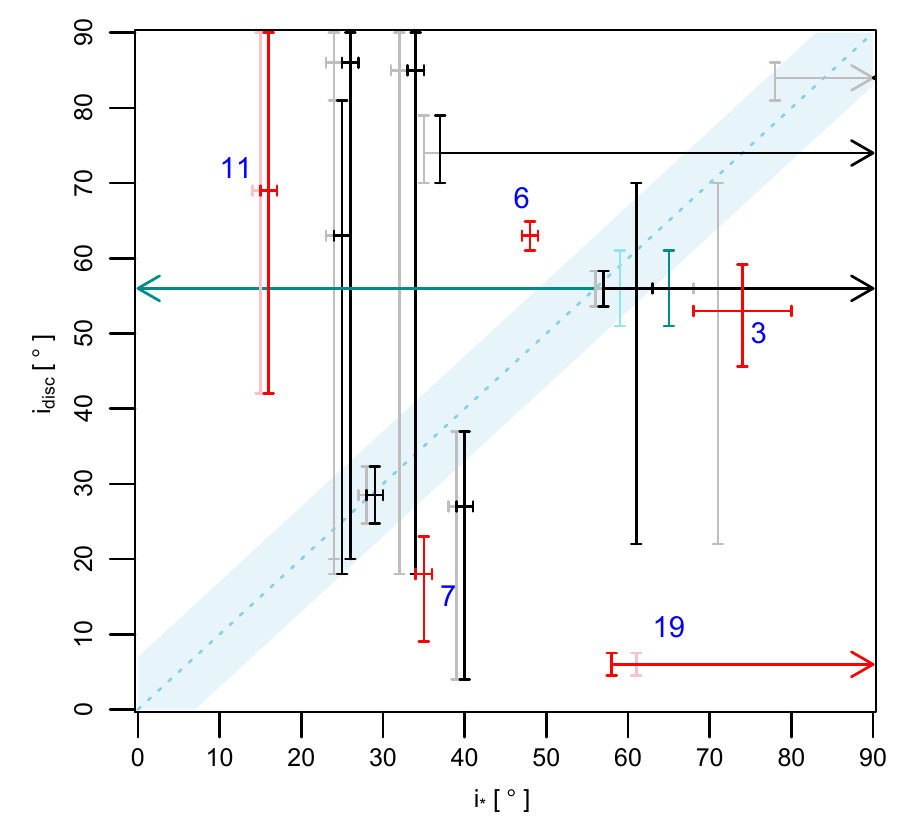}\hspace{-0.2cm}\vspace{-0.2cm}
    \caption{Disc and stellar inclinations, ($i_{\rm{disc}}$ and $i_{\star}$, respectively) for the 15 objects for which an estimate of $i_{\star}$ could be made. Black, red and green data points correspond to estimates of $i_{\star}$ made using \textit{Gaia} parallax measurements while data shown in grey, pink and cyan correspond to those using distances of $137.3\pm1.2\,$pc and $145\pm2\,$pc for $\rho$ Oph and Upper Sco members, respectively. The blue dashed line marks the location of $i_{\rm{disc}}= i_{\star}$ while the blue shaded region represents the scale of misalignment present in the solar system ($\pm7^{\circ}$; \citealt{Beck05}), typically considered to be an aligned system. Green and cyan colours are used for USco J160900.7-190852 (for which only an upper limit to $v\sin i$ is available) to aid distinguishing this object from GY~314 and USco~J160357.9-194210. The five systems with $i_{\rm{disc}}$ measurements which differ from their $i_{\star}$ estimates by $>1\sigma$ are highlighted in red and pink and labelled using the numerical identifiers in column 1 of Table~\ref{tab:lit_values}.}
    \label{fig:mis_ang}
\end{figure}

A comparison of the disc and stellar inclinations for the 15 systems in Table~\ref{tab:inc} with $i_{\star}$ estimates is presented in Fig.~\ref{fig:mis_ang}. In each case, both estimates of $i_{\star}$ are shown: \textit{Gaia} $DM$ estimates in black, red and green; SFR $DM$ estimates in grey, pink and cyan. Systems for which only lower limits to $i_{\star}$ could be estimated are marked by right-hand arrows. Green and cyan left-hand arrows are used as markers for USco~J160900.7-190852 (for which only an upper limit to $v\sin i$ was available in the literature) to help in distinguishing this object from the two other objects with the same disc inclination (GY~314 and USco~J160357.9-194210).

The dashed line in Fig.~\ref{fig:mis_ang} marks the location of star-disc alignment (i.e. $|i_{\star}-i_{\rm{disc}}|=0^{\circ}$) while the blue shaded area marks the $\pm7^{\circ}$ region either side of this line. This corresponds to the degree of misalignment present between the ecliptic plane and the solar equator \citep{Beck05}. Star-disc systems lying above or below the dashed line but within the shaded region could only be considered misaligned if one also considers the solar system to be so. Six of the ten systems with $i_{\star}$ estimates (not including those with lower limits) are consistent with $|i_{\star}-i_{\rm{disc}}|\leq 7^{\circ}$ and are thus considered to be aligned. For USco~J160900.0-190836, USco~J160959.4-180009, and USco~J155624.8-222555, this is certainly helped by the large lower bounds to their disc inclination uncertainties. For others, the alignment is more tightly constrained: Elias~2-24, for instance, has $i_{\star}=29\pm1^{\circ}$ (adopting \textit{Gaia} $DM=5.689$) while $i_{\rm{disc}}=28.5\pm3.8^{\circ}$ \citep{Dipierro18}.

For HIP~79977, the $\sin i$ estimate made using the \textit{Gaia} $DM$ also exceeds a value of $1$ by $>1\sigma$. However, by using the SFR $DM$, the uncertainty on $\sin i$ means that the estimated value is consistent with $i_{\star}\approx90^{\circ}$ within $1\sigma$. The inferred $1\sigma$ lower bound of $78^{\circ}$ for $i_{\star}$ using the SFR $DM$ is consistent with the relatively high disc inclination obtained for this object ($84^{+2}_{-3}$\,$^{\circ}$; \citealt{Thalmann13}). The fact that $\sin i$ exceeds a value of $1$ for this object is likely to be symptomatic of the increased measurement accuracy required to infer stellar inclinations close to $90^{\circ}$. $1\sigma$ lower limits to $i_{\star}$ are also provided in Table~\ref{tab:inc} for the remaining three systems with $\sin i$ estimates within $1\sigma$ of $\sin i = 1$ (GY~314, USco~J160823.2-193001, and RX~J1604.3-2130A). Each of these systems have high-precision disc inclination estimates. As a result, insights can still be drawn as to the degree of misalignment (if any) present in these systems. As Fig.~\ref{fig:mis_ang} clearly shows, both GY~314 and USco~J160823.2-193001 have stellar inclination lower limits consistent with their disc inclination measurements. In contrast, star-disc misalignment appears to be present in RX~J1604.3-2130A with $|i_{\star}-i_{\rm{disc}}|=52\pm32^{\circ}$ using \textit{Gaia} $DM=5.873$. This star (labelled system 19 in Fig.~\ref{fig:mis_ang}) is discussed further in Section~\ref{sec:2130284}. 

\begin{figure*}
    \centering
    \includegraphics[trim=0.0cm 0.0cm 0.0cm 0.0cm, clip=true, width=0.98\textwidth]{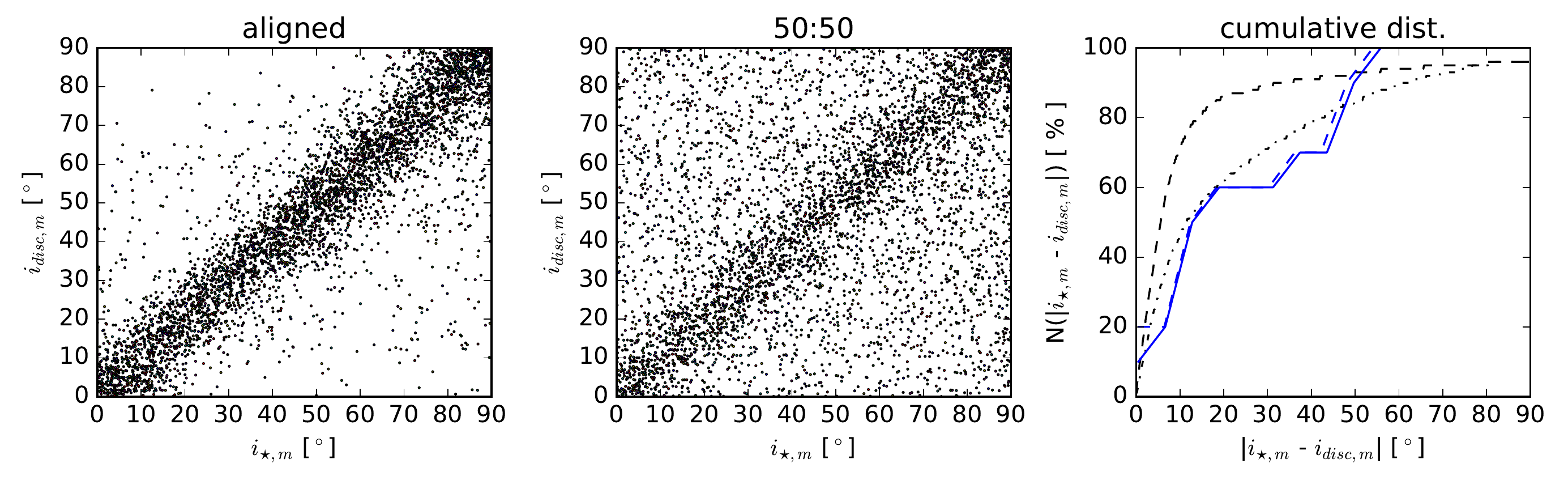}
    \caption{Distribution of simulated measurements of $i_{\rm{disc}}$ and $i_{\star}$ for a sample of 10,000 star-disc systems accounting for characteristic measurement uncertainties and projection effects. In the left-hand panel, $100\%$ of systems are aligned. In the central panel, $50\%$ of systems are aligned while $50\%$ are randomly oriented. In both cases, the true simulated star and disc inclinations vary from $-180^{\circ}$ to $180^{\circ}$ and are projected onto one $90^{\circ}$ quadrant, representing the observed inclination degeneracy (see Fig.~\ref{fig:sketch}). The right-hand panel compares the cumulative distributions of $|i_{\star,m}-i_{\rm{disc,m}}|$ for these two scenarios (black dashed and dot-dashed lines, respectively) to that observed in Fig.~\ref{fig:mis_ang} (solid and dashed blue lines represent the SFR $DM$ and \textit{Gaia} $DM$ $i_{\star}$ estimates, respectively). }
    \label{fig:distribution}
\end{figure*}

The four remaining systems, namely DoAr~24E, DoAr~25, SR~21A, and RX~J1603.9-2031A, each display $|i_{\star}-i_{\rm{disc}}|>7^{\circ}$ at $>1\sigma$. These are identified in Fig.~\ref{fig:mis_ang} by the numbers 3, 6, 7, and 11, respectively. The most statistically significant misalignment ($7.5\sigma$) is exhibited by DoAr~25 with $|i_{\star}-i_{\rm{disc}}|=15^{\circ}$. DoAr~24E appears misaligned at the $2.8\sigma$ level with $|i_{\star}-i_{\rm{disc}}|=21^{\circ}$ while SR~21A and RX~J1603.9-2031A both appear misaligned at the $2\sigma$ level ($|i_{\star}-i_{\rm{disc}}|=53^{\circ}$ and $17^{\circ}$, respectively). DoAr~24E and SR~21A exist in wide binary systems \citep{Chelli88, Reipurth93, Simon95} while RX~J1603.9-2031A, although close in the sky to the known triple system RX J1603.9-2031B ($50.4$'' sky-projected separation), appears to be single \citep{Kohler00, Lafreniere14}. DoAr~25 was included in surveys for multiplicity \citep{Costa00, Ratzka05, Cheetham15, Ruiz16, Kohn16} with no companion found to date. The possibility of prior dynamical interaction via a fly-by (e.g. \citealt{Cuello18}) cannot be ruled out. Further investigation of the possible cause of these apparent star-disc misalignments is beyond the scope of this paper.

\subsection{Impact of systematic $v\sin i$ uncertainties}\label{sec:systematics}
In Section~\ref{sec:sample}, the potential for systematic uncertainties being present among the $v\sin i$ measurements retrieved from the literature was raised. In particular, the measurements retrieved from \citet{Doppmann03} are potentially overestimated by up to $10\%$ as the spectral lines used in the measurement of $v\sin i$ are susceptible to Zeeman broadening effects. Thus, it is important to consider the potential impact this has on the results outlined above.

In particular, this has the potential to affect the inferred $i_{\star}$ estimates for DoAr~24E, Elias~2-24 and GY~314. If $v\sin i$ is overestimated by $10\%$, this would lead to a similar overestimation of $\sin i$. For Elias~2-24, this would lead to a reduction in $\sin i$ such that $i_{\star}=25\pm1^{\circ}$, maintaining the strong case for star-disc alignment in this system ($|i_{\star}-i_{\rm{disc}}|=3\pm3.8^{\circ}$). For DoAr~24E, the reduction in $\sin i$ reduces the value of $|i_{\star}-i_{\rm{disc}}|$ such that the object is consistent with the degree of misalignment present in the solar system and, as such, would not typically be interpreted as being misaligned ($|i_{\star}-i_{\rm{disc}}|=7^{\circ}$ at $1\sigma$ significance). Finally, the reduction in $v\sin i$ for GY~314 results in a value of $\sin i<1$ such that $i_{\star}=74\pm4^{\circ}$. This further implies that GY~314 features star-disc misalignment ($|i_{\star}-i_{\rm{disc}}|=18^{\circ}$ at $4.5\sigma$ significance). Taken together, this indicates no net impact in terms of the total number of systems which appear misaligned.

\subsection{Comparison to Monte Carlo simulations of test data}
To further assess whether the misalignments in Fig.~\ref{fig:mis_ang} can be considered significant given the measurement uncertainties, projection effects, and sample size, Monte Carlo simulations of two test populations were conducted. In the first simulated dataset, in which $100\%$ of the star-disc systems are aligned, a single value for the true inclination was drawn from a uniform distribution of possible inclinations between $-180$ and $180^{\circ}$ and assigned as both the true stellar inclination ($i_{\rm{\star,t}}$) and the true disc inclination ($i_{\rm{disc,t}}$). In the second simulated dataset, $i_{\rm{\star,t}}$ and $i_{\rm{disc,t}}$ were assigned in the same way for $50\%$ of the sample. For the other $50\%$, values of $i_{\rm{\star,t}}$ and $i_{\rm{disc,t}}$ were drawn separately from the uniform distribution of possible inclinations. In this way, $50\%$ of the sample are aligned while $50\%$ are randomly oriented.

The observed sample of disc and stellar inclinations shown in Fig.~\ref{fig:mis_ang} were then used to evaluate characteristic measurement uncertainties for $i_{\star}$ and $i_{\rm{disc}}$ to convert $i_{\rm{\star,t}}$ and $i_{\rm{disc,t}}$ to their respective ``observed'' values. From Table~\ref{tab:inc}, the typical uncertainty on $i_{\star}$ is roughly uniform at $1^{\circ}$ up to stellar inclinations of $40^{\circ}$. For $i_{\star}>40^{\circ}$, the uncertainty increases exponentially ($\approx e^{0.04i_{\star}}$). Simulated ``observed'' values of the stellar inclination ($i_{\rm{\star,o}}$) were drawn from a normal distribution of possible inclinations centred on $i_{\rm{\star,t}}$ with standard deviation given by the $i_{\star}$-dependent characteristic measurement uncertainty. No such trend with disc inclination was observed for the uncertainties on $i_{\rm{disc}}$ in Table~\ref{tab:lit_values}. Instead, the median lower and upper bounded uncertainties were used to construct an asymmetric probability distribution centred on $i_{\rm{disc,t}}$ with apparent standard deviation of $+5^{\circ}$ and $-9^{\circ}$. Simulated ``observed'' values of the disc inclination ($i_{\rm{disc,o}}$) were drawn from this distribution using the \textsc{rv discrete} function of the scipy.stats package in Python \citep{Jones01}.

Finally, the $i_{\rm{\star,o}}$ and $i_{\rm{disc,o}}$ values were corrected for degeneracies involved in measuring the inclination (see Fig.~\ref{fig:sketch}) such that all simulated measurements ($i_{\rm{\star,m}}$ and $i_{\rm{disc,m}}$) exist in the range $0-90^{\circ}$. In each simulated dataset, 10,000 measurements of the disc and stellar inclination were realised. The apparent dearth of data points in the top left corner of the left panel of Fig.~\ref{fig:distribution} is a real effect and is attributed to the asymmetric distribution of disc inclination measurement uncertainties used to generate this plot. The resulting $i_{\rm{\star,m}}$ and $i_{\rm{disc,m}}$ for the $100\%$ aligned and $50\%$ aligned; $50\%$ randomly oriented simulated datasets are presented in the left and middle panels of Fig.~\ref{fig:distribution}, respectively. The $100\%$ aligned test population indicates that $40\%$ of systems in our sample can be expected to display $|i_{\rm{\star,m}}-i_{\rm{disc,m}}|>7^{\circ}$ even if $|i_{\rm{\star,t}}-i_{\rm{disc,t}}|=0^{\circ}$ across the entire sample. By comparison, if $50\%$ of the star-disc systems have randomly oriented inclinations, the percentage of systems expected to display $|i_{\rm{\star,m}}-i_{\rm{disc,m}}|>7^{\circ}$ rises to $63\%$. Here, $40\%$ of the ten systems with $\sin i$ determinations indicate the presence of star-disc misalignment. If one widens this to include the four systems with lower limits to $i_{\star}$, the frequency of apparent star-disc misalignment in the sample reduces to $36\%$. Thus, it may be possible that the star-disc misalignments inferred for DoAr~24E, DoAr~25, SR~21A, RX~J1603.9-2031A, and RX~J1604.3-2130A merely arise as a result of measurement uncertainties and projection effects.

The right panel of Fig.~\ref{fig:distribution} compares the cumulative distribution of $|i_{\rm{\star,m}}-i_{\rm{disc,m}}|$ for the $100\%$ aligned (black dashed line) and $50\%$ aligned; $50\%$ randomly oriented (black dot-dashed line) simulated datasets to that of the ten systems with star and disc inclination estimates (blue lines). A double-sided Kolmogorov-Smirnov (KS) test indicates that the observed sample of $|i_{\rm{\star,m}}-i_{\rm{disc,m}}|$, calculated using the \textit{Gaia} $DM$ measurements, is consistent with being drawn from the same parent population as the $100\%$ aligned or the $50\%$ aligned; $50\%$ randomly oriented simulated populations at probabilities of $0.0020$ and $0.22$, respectively. Similar results are found when using the SFR $DM$ measurements: probabilities of $0.00026$ and $0.066$, respectively. Thus, although the frequency of apparent star-disc misalignments observed appears consistent with the simulated population of $100\%$ aligned systems (with characteristic measurement uncertainties and projection effects taken into account), the distribution of observed $|i_{\rm{\star,m}}-i_{\rm{disc,m}}|$ appears not to be. Instead, the observed distribution of $|i_{\rm{\star,m}}-i_{\rm{disc,m}}|$ is more consistent with the $50\%$ aligned; $50\%$ randomly oriented simulated population.

\subsection{RX J1604.3-2130A}\label{sec:2130284}
The lower limits to $i_{\star}$ determined for RX~J1604.3-2130A (see Table~\ref{tab:inc}) suggest that the star and the outer regions of this transitional disc are strongly misaligned at the $1.6\sigma$ level, with $|i_{\star}-i_{\rm{disc}}|>52^{\circ}$. As discussed previously, systematic errors can artificially increase $i_{\star}$ above its true value. Looking at this scenario from a different perspective, if one assumes that the star is aligned with the outer regions of its disc, the near pole-on geometry of the star would make measurement of $P$ difficult as the starspots present in a particular hemisphere of the star will not give rise to measurable rotation-modulated stellar flux variations unless they exist at very low latitudes. By comparison, Doppler imaging studies typically infer the presence of large polar spots on active stars like the PMS stars considered here \citep{Strassmeier02}. Furthermore, a star viewed pole-on would exhibit no rotational broadening of its photospheric spectral lines. Thus, the fact that $P$ and $v\sin i$ measurements exist for this object may be taken as evidence in itself that at least some degree of misalignment is present in the system. 

Interestingly, the LC of RX~J1604.3-2130A also exhibits short-period ``dipper'' events \citep{Rebull18, Ansdell16b}. The dipper phenomenon is generally attributed to occultation of the stellar photosphere by clumps of dusty disc material \citep{Ansdell16} but this has typically been considered to require disc inclinations of at least $\sim70^{\circ}$. As the outer disc regions of RX~J1604.3-2130A are observed close to face-on, they are unlikely to be responsible for the observed stellar occultation events. \citet{Ansdell16b} suggested the problem of observing a face-on outer disc and dipper LC phenomena could be reconciled via occultation of the stellar photosphere by accretion streams \citep{McGinnis15, Bodman17} if an inner disk warp exists. Indirect evidence for optically thick material existing closer to the star in a plane inclined with respect to the outer disc has been inferred from the presence of shadows in the outer disc \citep{Takami14, Pinilla18}. The inferred large misalignment angle between the inner and outer disk regions suggests that the stellar inclination estimated herein may be consistent with the inclination of the inner disc material even if the star and outer disc are misaligned.

The cause of such a star-outer disc misalignment remains unclear. \citet{Facchini14} have suggested that disc warps could be caused by embedded protoplanets on inclined orbits while \citet{Canovas17} argue that the structure of the transition disc of RX~J1604.3-2130A is most likely caused by the presence of multiple sub-stellar companions interior to the outer disc. Alternative scenarios for gap clearing which do not require the existence of protoplanets do exist but (i) the complex radial structure of the disc of RX~J1604.3-2130A makes gap carving in the dust disc via photoevaporation or grain growth unlikely and (ii) the additional existence of a gas cavity \citep{Zhang14} is difficult to reconcile with dead zone-induced dust gaps (e.g. \citealt{Ruge16}). Thus far, no companion within the disc of RX~J1604.3-2130A has been found. \citet{Kraus08} surveyed the object using aperture masking interferometry and direct imaging, ruling out companions down to $\Delta K=3.6\,$mag between $\sim10-20\,$mas of the central star, $\Delta K=5.4\,$mag between $\sim20-40\,$mas, and $\Delta K=6.2\,$mag between $\sim40-160\,$mas. High contrast SPHERE observations \citep{Canovas17} have also ruled out the existence of companions down to $2-3\,\rm{M_{Jup}}$ between $\sim0.15-0.8''$ while \citet{Ireland11} ruled out the existence of companions down to $\Delta K=3.4\,$mag between $\sim0.3-1''$.

Further from the star, a possible wide companion to RX~J1604.3-2130A at $16.22''$ was identified by \citet{Kraus09}. It remains to be seen whether this seemingly co-moving and co-distant companion (namely RX~J1604.3-2130B, see Table~\ref{tab:bin}) is close enough to be responsible for dynamically perturbing the disc. 

\section{Summary}\label{sec:concl}
In this study, I have evaluated the degree to which star-disc misalignments are present among young stellar objects in the $\rho$~Ophiuchus and Upper Scorpius SFRs. Due to observational limitations, this analysis is restricted to evaluating misalignments between the inclination angles of the stellar and disc angular momentum vectors. The data required to assess the degree of star-disc inclination misalignment exists in the literature for $20$ members of these two SFRs. However, the analysis presented in Section~\ref{sec:results} is limited to fifteen star-disc systems as (i) individual component photometry was not available for three close binary systems, (ii) the K2 photometry around ISO-Oph~51 and its $\sim9''$ companion appears confused (see Section~\ref{sec:multiplicity}), and (iii) the $\sin i$ estimate for WSB~63 exceeded a value of $1$ by $>1\sigma$. It was only possible to determine an upper limit to $i_{\star}$ for one of these fifteen systems while another four were limited to $1\sigma$ lower limit estimates of $i_{\star}$ as their $\sin i$ estimates exceeded a value of $1$ by $<1\sigma$.

An initial assessment of the degree of misalignment present in the sample was made by evaluating $|i_{\star}-i_{\rm{disc}}|$. Systems with $|i_{\star}-i_{\rm{disc}}|\geq7^{\circ}$, accounting for the measurement uncertainties on their $i_{\star}$ and $i_{\rm{disc}}$ estimates, were identified as showing evidence of potential star-disc misalignment. The value of $7^{\circ}$ reflects the degree of misalignment present in the solar system \citep{Beck05} which is typically considered to be an aligned system. Four out of the ten systems with $i_{\star}$ estimates appeared misaligned: DoAr~24E ($|i_{\star}-i_{\rm{disc}}|=21^{\circ}$ at $2.8\sigma$ significance); DoAr~25 ($|i_{\star}-i_{\rm{disc}}|=15^{\circ}$ at $7.5\sigma$ significance); SR~21A ($|i_{\star}-i_{\rm{disc}}|=53^{\circ}$ at $1.9\sigma$ significance); RX~J1603.9-2031A ($|i_{\star}-i_{\rm{disc}}|=17^{\circ}$ at $2.0\sigma$ significance). In addition, one out of the four systems with only $i_{\star}$ lower limits (RX~J1604-2130A) appeared misaligned: $|i_{\star}-i_{\rm{disc}}|>52^{\circ}$ ($1.6\sigma$ significance). In contrast, the strongest case for alignment between the star and disc inclinations is found for Elias~2-24 where $|i_{\star}-i_{\rm{disc}}|=0.5\pm3.8^{\circ}$.

The possible influence of systematic uncertainties introduced by the $v\sin i$ measurements retrieved from \citet{Doppmann03} were also investigated. These measurements utilised spectral lines which may be affected by Zemman broadening. As such, these $v\sin i$ measurements may be overestimated by as much as $10\%$. When taken into account, Elias~2-24 still appears consistent with star-disc alignment while the star-disc misalignment inferred for DoAr~24E reduces to levels consistent with that found in the solar system. However, the fraction of systems indicating the presence of misalignment remains consistent as the reduced $v\sin i$ value for GY~314 enables $i_{\star}$ to be calculated for this object, implying $|i_{\star}-i_{\rm{disc}}|=18^{\circ}$ ($4.5\sigma$ significance).

As shown in Fig.~\ref{fig:sketch}, comparing disc and stellar inclinations only probes one dimension of a three dimensional problem: a star with its south pole facing the observer is indistinguishable from one with its north pole tilted toward the observer while the position angle of the stellar rotation axis is not probed at all. A second assessment of the degree of misalignment present in the sample was made which accounted for the chance misalignment of truly aligned systems due to projection effects and characteristic measurement uncertainties on $i_{\star}$ and $i_{\rm{disc}}$. Monte Carlo simulations of two populations of test data were created. In one test population, $100\%$ of star and disc inclinations were equal while in the second test population, $50\%$ of the star and disc inclinations were randomly oriented. This test showed that the frequency of star-disc inclination misalignments observed in the sample (four out of ten) could be reproduced by the $100\%$ aligned population. However, KS tests indicated that the distribution of observed $|i_{\star}-i_{\rm{disc}}|$ was inconsistent with being drawn from the same parent population as the $100\%$ aligned test case.

The results presented here are in contrast to similar analyses performed for stars hosting warm debris discs \citep{Watson11, Greaves14}. The eighteen star-debris disc systems studied all showed preference for star-disc alignment through comparison of $i_{\star}$ and $i_{\rm{disc}}$. This result was used to suggest that the dynamical effects leading to the formation of high obliquity hot Jupiters do not operate on the planetesimals in the outer disc. The misalignments uncovered within the sample of PMS star-disc systems considered here suggest that this interpretation may not be entirely correct. Instead, the discrepancy between the degree of misalignment in protoplanetary and debris disc systems could hint at a possible link between star-disc-planet alignment and the formation and/or survival of debris discs. 

Further investigation into the fraction of star-disc misalignments among protoplanetary disc systems will be possible through the ongoing Gaia-ESO spectrographic survey \citep{Gilmore12}. Through its provision of high-resolution spectra for a large proportion of members of local star forming regions, consistent measurements of $v\sin i$ and $R_{\star}$ will be possible to determine for a larger fraction of $\rho$ Oph and Upper Sco members with K2 rotation periods and spatially resolved discs. 

\section*{Acknowledgements}
I thank the referee whose thorough and detailed review greatly improved the content of this paper. This work is supported by the ERC Starting Grant ``ImagePlanetFormDiscs'' (Grant Agreement No. 639889) and has made use of the SIMBAD database, operated at CDS, Strasbourg, France; the VizieR catalogue access tool, CDS, Strasbourg, France; and NASA's Astrophysics Data System. This work has made use of data from the European Space Agency (ESA) mission {\it Gaia} (\url{https://www.cosmos.esa.int/gaia}), processed by the {\it Gaia} Data Processing and Analysis Consortium (DPAC, \url{https://www.cosmos.esa.int/web/gaia/dpac/consortium}). Funding for the DPAC has been provided by national institutions, in particular the institutions participating in the {\it Gaia} Multilateral Agreement. Thanks go to Stuart
P. Littlefair, Jane S. Greaves, Kirstin Hay, Sam Morrell, and Tim Naylor for useful and interesting discussions. 




\bibliographystyle{mnras}
\bibliography{misalignment} 








\bsp	
\label{lastpage}
\end{document}